\documentclass[useAMS,usenatbib]{mn2e}

\usepackage{amssymb}
\usepackage{amsfonts}
\usepackage{amsmath}
\usepackage{graphicx,mathptm}
\usepackage{times}
\usepackage{natbib}

\def\kms{\ {\rm km\,s^{-1}}}
\def\lesssim{\mathrel{\hbox{\rlap{\hbox{\lower4pt\hbox{$\sim$}}}\hbox{$<$}}}}

\begin{document}

\title[QSO evolution at low redshifts]
{Modeling the QSO luminosity and
spatial clustering at low redshifts }

\author[F.~Marulli et al.]
  {F.~Marulli,$^1$ D.~Crociani,$^{1}$  M.~Volonteri,$^2$ 
  E.~Branchini$^3$  and L.~Moscardini$^1$ \\
  $^1$Dipartimento di Astronomia, Universit\`a degli Studi di Bologna,
      via Ranzani 1, I-40127 Bologna, Italy \\
  $^2$Institute of Astronomy,  Madingley Road, Cambridge, CB3 0HA, U.K. \\
  $^3$Dipartimento di Fisica, Universit\`a degli Studi ``Roma Tre'',
      via della Vasca Navale 84, I-00146 Roma, Italy\\
 }
\pagerange{\pageref{firstpage}--\pageref{lastpage}}
\pubyear{2005}

\label{firstpage}
 
\maketitle

\begin{abstract}

We investigate the ability of hierarchical models of QSO formation and
evolution to match the observed luminosity, number counts and spatial
clustering of quasars at redshift $z<2$. These models assume that the
QSO emission is triggered by galaxy mergers, that the mass of the
central black hole correlates with halo properties and that quasars
shine at their Eddington luminosity except, perhaps, during the very
early stages of evolution. We find that models based on simple
analytic approximations successfully reproduce the observed $B$-band
QSO luminosity function at all redshifts, provided that some
mechanisms is advocated to quench mass accretion within haloes larger
than $\sim 10^{13} M_{\odot}$ that host bright quasars. These models also
match the observed strength of QSO clustering at $z \sim 0.8$. At
larger redshifts, however, they underpredict the QSO biasing which,
instead, is correctly reproduced by semi-analytic models in which the
halo merger history and associated BHs are followed by Monte Carlo
realizations of the merger hierarchy.
We show that the disagreement
between the luminosity function predicted by semi-analytic models and
observations can be ascribed to the use of B-band data, which are a
biased tracer of the quasar population, due to obscuration.

\end{abstract}

\section{introduction}

Observational evidences accumulated over the past few years indicate
that the physical properties of supermassive black holes [SBHs,
hereafter] residing at the centre of most, if not all, spheroidal
galaxies \citep{richstone1998}, correlate with those of the host
galaxy \citep{magorrian1998, ferrarese2000, gebhardt2000} and,
possibly, of the host dark matter (DM) halo \citep{ferrarese2002}.
Although it is not clear whether the observed correlation between halo
and BH masses is genuine or simply reflects the fact that massive
haloes preferentially host massive spheroids \citep{wyithe2005a}, the
strong connection between the BH properties and the gravitational
potential wells that host them suggests a link between the assembly of
BHs and the evolution of galaxy spheroids. This co-evolution is
expected within the frame of hierarchical models of structure
formation, like the popular cold dark matter (CDM) cosmogonies. Since
it is generally accepted that the quasar activity is powered by SBHs,
several models have been proposed to trace the assembly of SBHs during
the hierarchical build-up of their host haloes using numerical,
analytic and semi-analytic methods. In this work we focus on the
latter two approaches that consist of parameterizing in term of simple
analytic models the complex physics of the galaxy formation process,
while the evolution of DM haloes can be either followed using
numerical techniques, in the so called semi-analytic approach, or
simplified analytic prescriptions, which constitute a full analytic
model.

Over the years many different analytic \citep[see,
e.g.,][]{efstathiou1988, haehnelt1993, haiman1998, percival1999, haiman2000, 
martini2001, hatziminaoglou2001, wyithe2002, wyithe2003, hatziminaoglou2003} 
and semi-analytic \citep[see, e.g.,][]{Cattaneo1999, kauffmann2000, cavaliere2000, 
cattaneo2001, cavaliere2002, enoki2003, volonteri2003a, springel2005, 
cattaneo2005} hierarchical models have been proposed
to understand the QSO phenomenon. Most of these models successfully
reproduce the QSO luminosity function [LF] at high (i.e. $z> 2$)
redshifts, when the QSO activity reaches its peak, and some of them
also account for the existence of QSO at very high (i.e.. $z\sim 6$)
redshift, hosting BHs as massive as $10^{9}M_{\odot}$
\citep{volonteri_rees2005}. However, hierarchical models generally
fail to match the QSO properties at lower redshifts since the halo
merger rate, which is generally regarded as the mechanism that
triggers the QSO activity, declines with time less rapidly than the
observed QSO number density. Various possible solutions have been
proposed to solve this problem. For example it has been suggested
that encounters within galaxy groups and clusters play a fundamental
role in triggering the BH accretion \citep{cavaliere2000},
that either the fraction of accreted gas or the time-scale of accretion
depend on redshift \citep{haiman2000, cattaneo2001}, or that the accretion
rate is proportional to the gas density of the host galaxy \citep{cattaneo2005}.
These solutions improve, with a different degree of success,
the match to the observed LF at low redshift. The model proposed
by Enoki et al. 2003 also allows to reproduce the QSO clustering,
but is not guaranteed to match the correlation of Ferrarese (2002)
between BH and halo mass.
Another possibility, suggested by the outcome of high resolution,
hydrodynamical simulations \citep{hopkins2005}, is that of assuming
that for a large fraction of its lifetime the QSO is heavily obscured
by surrounding gas and dust except for a short window time of $\sim
10^{7}$ yr during which it shines at its peak luminosity.  

In this work, we investigate to which extent `standard' analytic and
semi-analytic hierarchical models, where the assembly of supermassive black holes 
is related to the merger history of DM haloes, can
reproduce both the observed QSO LF and their clustering at low
redshifts. For this purpose we implement the two original analytic
models of \cite{wyithe2002} (WL02 hereafter) and \cite{wyithe2003}
(WL03 hereafter) and the semi-analytic one of Volonteri, Haardt \&
Madau (2003) (VHM hereafter), compare their predictions to the LF and
clustering of the optical QSOs measured in various redshift catalogues,
and discuss possible modifications to the models that improve the
match to the data.

The outline of the paper is as follows. In Section 2 we present the
observational datasets and the QSO properties relevant for our
analysis. In Section 3 we present the WL02 and WL03 models, compare
their predictions with observations and introduce some modification to
the original models to better match the observed LF. A similar
analysis is repeated in Section 4 for the semi-analytic model of VHM.
Finally, in the last section we discuss our results and draw our main
conclusions.

Throughout this paper we assume a flat $\Lambda$CDM cosmological model
with Hubble constant $h=H_0/100 \kms {\rm Mpc}^{-1}=0.7$ and a
dominant contribution to the density parameter from the cosmological
constant, $\Omega_{\Lambda}=0.7$. We adopt a CDM density power
spectrum with primordial spectral index $n=1$ and normalized by
assuming $\sigma_8=0.9$.

\section{Datasets}
\label{sec:data}

The main dataset considered in this work is the QSO redshift
catalogue by \cite{croom2004} [C04 hereafter] obtained by merging the
2dF QSO Redshift Survey (2QZ), containing objects with an apparent
$b_j$ magnitude $18.25<b_j<20.85$, with the 6dF QSO Redshift Survey
(6QZ) of bright ($16<b_j<18.25$) quasars. The full sample includes
23660 quasars in a wide redshift range ($0.3 < z < 2.9$) and spread
over 721.6 deg$^2$ on the sky. The 2QZ/6QZ catalogue is affected by
various types of incompleteness described in details by C04 that need
to be accounted for in order to minimize systematic effects. For this
purpose we have considered various subcatalogues, described below,
characterized by a higher degree of completeness to allow a precise
measurement of the QSO luminosity function [LF] and clustering
properties.

A second dataset consisting of 5645 quasars with an apparent magnitude
$18.0<g<21.85$ extracted from the recent 2dF-SDSS LRG and QSO [2SLAQ]
survey \citep[][ R05 hereafter]{richards2005} has also been considered
to provide us with a reliable estimate of the LF of fainter objects at
$z<2.1$.
 
\subsection{The QSO Optical Luminosity Functions}
\label{sec:datalf}

\begin{figure*}
\includegraphics[width=0.96\textwidth]{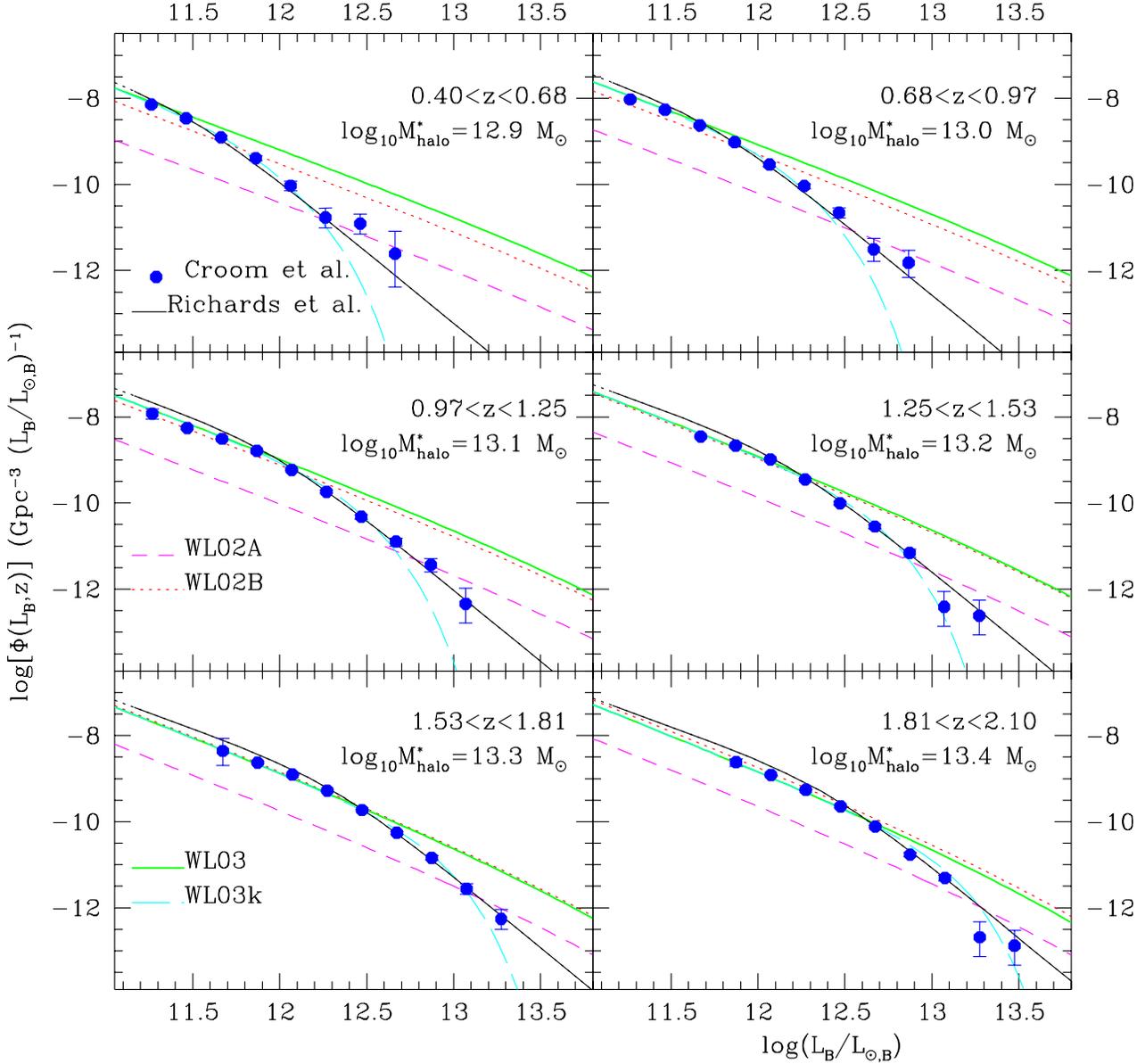}
\caption{
The QSO luminosity function in $B$-band at six different redshifts:
models vs. observations. The filled circles show the 2dF/6dF QSO
luminosity function measured by C04 together with their $1\sigma$
error bars. The thin solid line shows the best-fit to the 2SLAQ QSO
luminosity function of R05. The short-dashed and the dotted lines show
the WL02 model predictions obtained by setting $(\gamma=4.71, \
\epsilon_{0}=10^{-5.1}, \ t_{dc,0}=10^{6.3} {\rm yr})$ (label: WL02A)
and $(\gamma=4.71, \ \epsilon_{0}=10^{-5.1}, \ t_{dc,0}=10^{7.2} {\rm
yr})$ (label: WL02B), respectively. The thick solid line shows the
WL03 model predictions obtained for $(\gamma=5.0, \
\epsilon_{0}=10^{-5.7})$, while the thin long-dashed line (label:
WL03k) represents the same model with an exponential cut in the
luminosity-mass relation as in eq.(\ref{eq:lm}), with $k=0.77$ and
$M^{*}_{\rm halo}$ indicated in the plots.}
\label{fig:lf}
\end{figure*}

The 2QZ/6QZ catalogue has been used by C04 to compute the optical LF
of a subsample of 15830 QSOs brighter than $M_{b_j} = 22.5$ in the
redshift range $0.4<z<2.1$. The cut in absolute magnitude guarantees a
minimum spectroscopic sector completeness of at least 70 per cent,
while redshift constraints ensure a photometric completeness of 85 per
cent. The LF has been evaluated into $\delta M_{bj}=0.5$ bins in
absolute magnitude using the $1/V$ estimator of \cite{page2000} into
six equally spaced, independent redshift bins. The filled dots in the
six panels of Fig.~\ref{fig:lf} show the LF of C04 measured in the
different redshift intervals indicated in each plot.

In the same figure the thin solid line represents the best fit
to the 2SLAQ LF determined by R05 in the same redshift intervals and
$b_j$ magnitude bins. The original $g$ magnitudes have been
transformed into $b_j$ magnitudes by using the relation $\langle g -
b_j \rangle =-0.045$ of R05. Finally, $b_j$ magnitudes have been
converted into $B$-band magnitudes by using the relation
$M_B=M_{bj}+0.07$ of \cite{porciani2004} [PMN hereafter].

\subsection{QSO vs. Dark Matter Clustering}
\label{sec:databias}

PMN have estimated the QSO two-point spatial correlation function of
$\sim 14000$ 2QZ/6QZ quasars with redshift $0.8<z<2.1$ in three
different redshift intervals $[0.8,1.3]$, $[1.3,1.7]$ and $[1.7,2.1]$.
The three subsamples with median redshifts $z_{\rm eff}=1.06, 1.51,
1.89$ contain $\sim 4300$, $\sim 4700$ and $\sim 4900$ objects each.
The more conservative cut in redshift and the use of these three
redshift intervals guarantee (i) a photometric completeness larger
than 90 per cent, (ii) a similar number of quasars in each redshift bin
and (iii) that each subsample covers a similar interval of cosmic
time.

\begin{figure}
\includegraphics[width=0.45\textwidth]{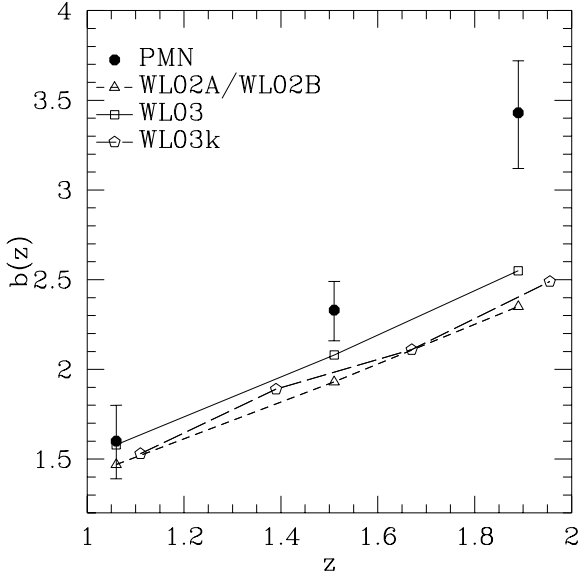}
\caption{
The mean QSO-to-mass biasing parameter, $b(z_{\rm eff})$ estimated at three
effective redshifts $z=1.06$, $z=1.51$ and $z=1.89$: models
vs. observations. The filled circles show the mean biasing of 2dF/6dF
quasars measured by PMN with the associated $1 \sigma$ uncertainties.
The short-dashed line shows the theoretical predictions
of the WL02 model with $(\gamma=4.71, \ \epsilon_{0}=10^{-5.1}, \
t_{dc,0}=10^{6.3} {\rm yr})$ (label: WL02A) and with $(\gamma=4.71, \
\epsilon_{0}=10^{-5.1}, \ t_{dc,0}=10^{7.2} {\rm yr})$ (label: WL02B).
The amount and evolution of QSO biasing of these two models are 
exactly the same. The thick solid line shows the WL03 model predictions
obtained for $(\gamma=5.0, \ \epsilon_{0}=10^{-5.7})$ and the thin
long-dashed line (label: WL03k) represents the same model with an
exponential cut in the luminosity-mass relation as in
eq.(\ref{eq:lm}), with $k=0.77$ and $M^{*}_{\rm halo}$ indicated in
the plot.}
\label{fig:bias}
\end{figure}

PMN have measured the spatial two-point correlation function,
$\xi^{\rm obs}(r_{\perp},\pi)$, using all QSO pairs separated by $\pi$
along the line of sight and by $r_{\perp}$ along the direction
perpendicular to it and have computed the `projected correlation
function' $\Xi$ by integrating $\xi^{\rm obs}$ along the $\pi$ direction:
\begin{equation}
\frac{\Xi^{\rm obs}(r_\perp)}{r_\perp}=\frac{2}{r_\perp} 
\int_0^{r_\perp} \xi^{\rm obs}(r_\perp,\pi) d\pi\ .
\label{eq:corrP}
\end{equation}
Then, in order to quantify the QSO clustering with respect to the
mass, they have evaluated the quasar-to-mass biasing function
\begin{equation}
b(r_\perp,z_{\rm eff})=\left[\frac{\Xi^{\rm obs}(r_\perp,z_{\rm min}<z<z_{\rm 
max})}{\Xi_{\rm DM}(r_\perp,z_{\rm eff})} \right]\ ,
\label{eq:biasP}
\end{equation}
where $z_{\rm min}$ and $z_{\rm max}$ represent the lower and upper
limits of the redshift interval and the projected mass correlation
function, $\Xi_{\rm DM}(r_{\perp})$, was estimated as in
\cite{peacock1996}. PMN have found that the QSO biasing function in
Eq.(\ref{eq:biasP}) is almost independent of the projected separation
$r_{\perp}$ and thus it is possible to characterize the QSO spatial
correlation properties at $z_{\rm eff}$ using a single `bias'
parameter $b(z_{\rm eff})$. The results of the PMN analysis are
presented in Fig.~\ref{fig:bias}, where the filled dots show the value
of $b(z_{\rm eff})$ for the 2dF/6dF QSOs at three different redshifts,
together with their $1\sigma$ uncertainty. The QSO-to-mass bias
parameter $b(z_{\rm eff})$ increases with redshift, in quantitative
agreement with the results of a similar analysis performed by
\cite{croom2005}.

\section{Analytic models for the QSO evolution}
\label{sec:model_an}

In this section we review the two analytic models by WL02 and WL03, use
them to predict the QSO LF and clustering at $z<2$ and compare the
results to the observations described in Sections \ref{sec:datalf} and
\ref{sec:databias}. Since we apply these models beyond the redshift
range they were originally designed for, we discuss and introduce some
modifications to improve the fit to the 2dF/6DF and 2SLAQ LFs and
clustering.

\subsection{The WL02 model}
\label{sec:model_wl02}

The WL02 model describes the QSO evolution within the theoretical
framework of hierarchical structure formation. However, instead of
associating the QSO activity to the formation of DM virialized haloes,
as originally proposed by \cite{haiman1998}, it assumes that the QSO
phenomenon is triggered by halo-halo mergers. The model also assumes
that the mass of the BH powering the QSO, $M_{\rm bh}$, is a
fraction, $\epsilon$, of the host halo mass $M_{\rm halo}$
and that, after a merging event, the QSO shines at the Eddington
luminosity, $L_{\rm Edd}$, with an universal light curve, $f(t)$. The
$B$-band QSO luminosity can be related to $M_{\rm bh}$ and $M_{\rm
halo}$ through $f(t)$:
\begin{equation}
L_{B}(t)=M_{\rm bh}f(t)=\epsilon M_{\rm halo}f(t) \ \ \ \ 
{\rm for} \ \ \ \ M_{\rm halo}> M_{\rm min}\ ,
\label{eq:corr}
\end{equation}
where $M_{\rm min}\sim 10^{8}[(1+z)/10]^{-3/2} M_{\odot}$ is the
minimum halo mass inside which a BH can form. The number density of
active QSOs then can be obtained by multiplying the number of haloes
with mass between $\Delta M_{\rm halo}$ and $\Delta M_{\rm
halo}+d\Delta M_{\rm halo}$ that accrete onto a halo of mass $M_{\rm
halo}-\Delta M_{\rm halo}$ per unit time by the number density of
haloes in the same mass range:
\begin{multline}
I(M_{\rm halo},\Delta M_{\rm halo})\equiv\frac{dN(M,z)}{dM}\Big
\vert_{M=M_{\rm halo}-\Delta M_{\rm halo}}\\
   \times\frac{d^{2}N_{\rm merge}}{d\Delta M_{\rm halo}dt}\Big
\vert_{M=M_{\rm halo}-\Delta M_{\rm halo}}\ .
\end{multline}
The quantity $\frac{d^{2}N_{\rm merge}}{d\Delta M_{\rm halo}dt}$
represents the merging rate predicted by the Extended Press-Schechter
theory [EPS] \cite{lacey1993} and the halo mass function
$\frac{dN(M,z)}{dM}$ is that of \cite{sheth1999}.

A further relation between $M_{\rm bh}$ and $M_{\rm halo}$ can be imposed by
assuming that the scaling relation between the BH mass and the
circular velocity of the host DM halo, $v_{c}$, observed by
\cite{ferrarese2002} in the local universe, holds at all redshifts:
\begin{multline}
M_{\rm bh}\propto v_{c}^{\gamma}=(159.4)^{\gamma}
\Big(\frac{M_{\rm halo}}{10^{12}h^{-1}M_{\odot}}\Big)^{\gamma/3}\\
\times\Big(\frac{\Omega_{\rm m}(0)}{\Omega_{\rm m}(z)}\frac{\Delta_{c}}
{18\pi^{2}}\Big)^{\gamma/6}(1+z)^{\gamma/2}\ ,
\label{eq:mbhcv}
\end{multline}
where the second equality follows from the relation between $v_{c}$
and $M_{\rm halo}$ \citep{barkana2001} in which
$\Delta_{c}(z)=18\pi^{2}+82d-39d^{2}$, $d\equiv \Omega_{\rm
m}(z)-1$ and $\Omega_{\rm m}(z)$ represents the matter density
parameter at a given redshift $z$. From eqs.(\ref{eq:corr}) and 
(\ref{eq:mbhcv}):
\begin{equation}
\epsilon=\epsilon_{0}\Big(\frac{M_{\rm halo}}
{10^{12}M_{\odot}}\Big)^{\gamma/3-1}
\Big(\frac{\Omega_{\rm m}(0)}{\Omega_{\rm m}(z)}\frac{\Delta_{c}}
{18\pi^{2}}\Big)^{\gamma/6} h^{\gamma/3}(1+z)^{\gamma/2}\ .
\label{eq:eps}
\end{equation}

Finally, we assume that the QSO luminosity curve is given by a simple
step function
\begin{equation}
f(t)=\frac{L_{{\rm Edd},B}}{M_{\rm bh}} \theta \Big(\frac{\Delta
M_{\rm halo}} {M_{\rm halo}}t_{dc,0}-t\Big)\ ,
\label{eq:light}
\end{equation}
where $t_{dc,0} \ll H^{-1}(z) $ is the time of QSO duty cycle at
$z=0$.

Eqs.(\ref{eq:mbhcv}), (\ref{eq:eps}) and (\ref{eq:light}) allow us to
compute the QSO LF:
\begin{multline}
\Phi(L_{B},z)=\int_{\epsilon M_{\rm min}}^{\infty}dM_{\rm bh}
\int_{0}^{0.5\epsilon M_{\rm halo}}d\Delta M_{\rm bh}\\
\times\int_{z}^{\infty}dz^{\prime}\frac{dN_{\rm bh}}{dM}\Big
\vert_{M=M_{\rm bh}-\Delta M_{\rm bh}}
\frac{d^{2}N_{\rm merge}}{d\Delta M_{\rm bh}dt}\Big
\vert_{M=M_{\rm bh}-\Delta M_{\rm bh}}\\
\times\frac{dt^{\prime}}{dz^{\prime}}\delta [L_{B}-M_{\rm bh}f(t_{z}-t^{\prime})]
\label{eq:lumf1}
\end{multline}
that, once integrated over $M_{\rm bh}$, has an analytic expression that
depends on the three free parameters $t_{dc,0}$, $\epsilon_{0}$ and
$\gamma$:
\begin{multline}
\Phi(L_{B},z)=\int_{0}^{0.5M_{\rm halo}}d\Delta M_{\rm halo}
\frac{3}{\gamma\epsilon}\frac{t_{dc,0}}{5.7\times 10^{3}}
\frac{\Delta M_{\rm halo}}{M_{\rm halo}}\\
\times I(M_{\rm halo},\Delta M_{\rm halo})\ ,
\label{eq:lumf2}
\end{multline}
where $M_{\rm halo}=L_{{\rm Edd},B}/(5.7\times 10^{3}\epsilon L_{\odot,B})M_{\odot}$.

The connection between QSO luminosity and halo mass in our model,
$L_B(M_{\rm halo})$, and the existence of analytic models that
describe the spatial and clustering of DM haloes \citep[see,
e.g.,][]{mo1996, sheth1999} allow us to investigate the spatial
correlation properties for QSOs. In particular, we can compute the
QSO-to-mass biasing parameter, $b(z)$, averaged over all QSO
luminosities:
\begin{equation}
b(z)=\frac{\displaystyle\int_{L_{B,{\rm min}}}^{\infty}
b(L_B(M_{\rm halo}),z)
\Phi(L_B,z)dL_B}
{\displaystyle\int_{L_{{\rm min},B}}^{\infty}\Phi(L_B,z)dL_B}\ ,
\label{eq:bias}
\end{equation}
where $\Phi(L_B,z)$ is the QSO LF in Eq.(\ref{eq:lumf2}), $L_{B,{\rm
min}}$ is the luminosity of the faintest object in the sample. The
quantity $b(L_B(M_{\rm halo}),z)$ represents the biasing parameter of
a halo of mass $M_{\rm halo}$ hosting a QSO of luminosity $L_B$ at the
redshift $z$ that has been computed by \cite{sheth1999}:
\begin{multline}
b(L_B(M_{\rm halo}),z)=1+\frac{1}{\delta_c(0)}\Big[\frac{a\delta_{c}^{2}(z)}
{\sigma_{M}^{2}}-1\Big]\\
+\frac{2p}{\delta_{c}(0)}\Big(\frac{1}{1+[\sqrt{a}
\delta_{c}(z)/\sigma_{M}]^{2p}}\Big)\ ,
\label{eq:biasmodel}
\end{multline}
where $a=0.707$, $p=0.3$, $\delta_c(z)$ is the critical threshold on
the linear overdensity for spherical collapse at redshift $z$ and
$\sigma_{M}^2$ is the rms linear density variance smoothed with a
`top-hat' filter corresponding to the mass $M$.
Eq.(\ref{eq:bias}) provides us with an analytic expression for $b(z)$. 
It assumes a univocal relation between quasar luminosity and halo
mass, or, in other words, that the probability of finding a
quasar of a given luminosity $L_B$ in a halo of mass $M_{\rm halo}$
depends on $M_{\rm halo}$ only.

\subsection{WL02 model vs. observations}
\label{sec:wl02_lf}

Let us now compare both the QSO LF and the biasing parameter predicted
by the WL02 model to the various datasets available. Given the
background cosmology, the model predictions are fully specified by a
set of three parameters: $(\gamma, \ \epsilon_{0}, \ t_{dc,0})$. Here
we explore two separate cases. The first one, which is slightly
different from the one considered in the original WL02 paper (and labeled
WL02A in Figs.~\ref{fig:lf} and \ref{fig:bias}), uses the parameters
$(\gamma =4.71, \ \epsilon_{0}=10^{-5.1}, \ t_{dc,0}=10^{6.3}\ {\rm
yr})$. The first two chosen values represent the best fit to the
observations of \cite{ferrarese2002}, while the latter, which
corresponds to a value of $t_{dc}(z=3)=10^{6.9}$ yr at $z\sim 3$, is
fully consistent with the lifetime of bright QSOs, $t_{dc}(z \simeq
3)=10^{7}$ yr, inferred from the sample of Lyman-break galaxies of
\cite{steidel2002}. The predicted QSO LFs, plotted as short-dashed line in
Fig.~\ref{fig:lf}, fail to match the observed LF both in the bright
and the faint ends. This result is similar to the one originally obtained
by WL02 using the set of parameters $(\gamma =5.0, \
\epsilon_{0}=10^{-5.4}, \ t_{dc,0}=10^{6.3}\ {\rm  yr})$, that 
constitutes their best fit to the data. Since the overall amplitude
of the model LF (Eq. \ref{eq:lumf2}) linearly depends on
$\gamma^{-1}$, $\epsilon_{0}^{-1}$ and $t_{dc,0}$, it is quite
straightforward to boost up the model LF to match the number density
of the observed QSOs. For example, fixing the values of $\gamma
=4.71$, $\epsilon_{0}=10^{-5.1}$, and leaving $t_{dc,0}$ as a free
parameter, we find a best fit for $t_{dc,0}=10^{7.2}$ yr. The
resulting model, labeled WL02B, is shown in Fig.~\ref{fig:lf} as a
dotted line. This duty-cycle is very large and some {\it ad hoc}
modifications to the WL02 model would be required to satisfy the
high-redshift constraints of \cite{steidel2002}. We will discuss
physically motivated modifications to the WL02 model later on in the
framework of the WL03 model.

The biasing functions predicted by the WL02A and WL02B models are
shown in Fig.~\ref{fig:bias} at three different redshifts and compared
to the PMN data. The line-styles are the same as in
Fig.~\ref{fig:lf}. The two models predict the same amount and
evolution of QSO biasing. In both cases at $z=1.06$, QSOs are mildly
biased with respect to the underlying mass, marginally consistent with
observations, while their clustering at $z=1.89$ is significantly less
than observed.

\subsection{The WL03 model}
\label{sec:model_wl03}

WL03 modified the original WL02 model by prescribing a self-regulated
accretion mechanism of SBHs following {\it major} mergers
between haloes of different masses, as proposed by
\cite{kauffmann2000}. A self-regulated accretion mechanism is the
natural outcome of the production of powerful gas winds that interrupt
the infall of gas on the BH after halo mergers. Self-regulation takes
place when the energy in the outflow equals the gravitational binding
energy in a dynamical time \citep{silk1998}. Assuming that the gas is
located in a disk with characteristic radius $\sim 0.035r_{\rm vir}$,
the dynamical time $t_{\rm dyn}$ is given by
\begin{multline}
t_{\rm dyn}=0.035\frac{r_{\rm vir}}{v_{c}}=\frac{3.64\times 10^{7}}{h}
\Bigg(\frac{{\Omega_{\rm m}(0)}}{{\Omega_{\rm m}(z)}}
\frac{\Delta_{c}}{18\pi^{2}}\Bigg)^{-1/2}\\
\times(1+z)^{-3/2}\textrm{yr}\ ,
\end{multline}
and represents the quasar duty cycle: $t_{dc}=t_{\rm dyn}$. The major
merger condition has been introduced to guarantee that the dynamical
friction time-scale \citep{binney1987} for the satellite is shorter
than a Hubble time. For this reason WL03 only considered mergers
between haloes with a mass ratio $P\equiv\Delta M_{\rm halo}/M_{\rm
halo}>0.25$. As a consequence, the model LF can be expressed as
\begin{multline}
\Phi(L_{B},z)=\int_{0.25M_{\rm halo}}^{0.5M_{\rm halo}}d\Delta 
M_{\rm halo}\frac{3}
{\gamma\epsilon}
\frac{t_{\rm dyn}}{5.7\times 10^{3}}\\
\times I(M_{\rm halo},\Delta M_{\rm halo})\ .
\label{eq:lumfw03}
\end{multline}
It only depends on two free parameters: $\epsilon_{0}$ and $\gamma$.
The model biasing function has been computed using Eq.(\ref{eq:bias}).

\subsection{WL03 model vs. observations}
\label{sec:wl03_lf}

The first model we have explored is the one by WL03 corresponding to
the choice of parameters $(\gamma=5.0,\epsilon_{0}=10^{-5.7})$, still
consistent with the observational data of \cite{ferrarese2002}. The
model LF (labeled WL03 and plotted with a thick solid line in
Fig.~\ref{fig:lf}) is very similar to that of WL02B, but has the
advantage of having a physically, well motivated QSO duty-cycle. The
WL03 LF matches the observed one at low luminosities, but it
overpredicts the number density of bright QSOs. This discrepancy is
more evident at low redshifts and can be accounted for by modifying
the WL03 model. One possibility is to allow for a major merger
threshold $P$ that depends on $M_{\rm halo}$. As we have checked, it
is indeed possible to find some suitable function $P(M_{\rm halo})$
monotonically increasing with $M_{\rm halo}$ that allows to match both
the faint and bright ends of the observed QSO LF. While this is
somewhat an {\it ad hoc} solution, a more physically plausible
modification has been proposed by WL03 and consists of assuming that
accretion onto BH is hampered by the high temperature of the gas
within group/cluster-size haloes. WL03 proposed that accretion onto
the central BH should be prevented within haloes of masses larger than
$10^{13.5} M_{\odot}$, corresponding to a $L\sim 2\times 10^{13}
L_{\odot,B}$ at $z\simeq 1$. We have made a similar assumption and
proposed that the accretion efficiency decreases above a given
critical halo mass resulting in a modified relation between $L_{QSO}$
and $M_{\rm halo}$:
\begin{equation}\label{eq:lm}
L_{QSO}=\tilde{L}(1-\exp(-(L^{*}/\tilde{L})^{k}))\ .
\end{equation}
In the previous equation $\tilde{L}=5.7\times 10^{3}\epsilon(M_{\rm
halo}/M_{\odot})$ is the $B$-band Eddington luminosity 
of the original WL03 model and $L^{*}$ is the Eddington luminosity of a halo with
critical mass $M^{*}_{\rm halo}$. Both $k$ and $M^{*}_{\rm halo}$ can
be treated as free parameters. The thin dashed line in
Fig.~\ref{fig:lf} shows the effect of a keeping $k=0.77$ while leaving
$M^{*}_{halo}$ as a free parameter, whose value is indicated in the
plots. The LF predicted by the model (labeled WL03k) provides a good
match to the observed one at all redshifts, except perhaps for very
bright objects. The resulting values of $M^{*}_{\rm halo}$ range
between $10^{12.9}$ and $10^{13.4} M_\odot$ which constitute plausible
values, close to that proposed by WL03. Leaving also $k$ as a free
parameter does not improve significantly the agreement with the
observational data and a best fit value close to $k=0.77$ is found at
all redshifts.
We have also tried to use the model WL03 
without the restriction to major mergers, but it has resulted 
inconsistent with the data by about several orders of magnitude and we have 
decided not to show it.

\section{Semi-analytic models for the hierarchical evolution of quasars}
\label{sec:model_sa}

Semi-analytic approaches to the evolution of QSOs in a hierarchical
scenario also assume some relation between QSOs and DM haloes
properties. However, they differ from analytic modeling since the
evolution of DM haloes and of the QSOs within them are treated
separately. The merging history of DM haloes is described by the EPS
formalism. Phenomenological relations are used to model the physical
processes leading to the quasar evolution. It is therefore possible to
adopt a more detailed description of the physics involving the
baryonic component of cosmic structures, including BHs. A second
advantage of the semi-analytic approach is the flexibility of the
scheme, so that different astrophysical prescriptions can be tested
within the same framework of cosmic evolution of DM haloes.

\subsection{The VHM model}
\label{sec:model_vhm}

In this work we focus on the semi-analytic model developed by VHM that
describes the hierarchical assembly, evolution and dynamics of the BHs
powering QSOs (\citealt*{volonteri2003a, volonteri2003b}; \citealt{madau2004,  
volonteri_rees2005}). Like WL02 and WL03, VHM also assume that: $(i)$ the
observed correlation between BH masses and circular velocity
\citep{ferrarese2002} justifies the assumption of a link between QSO
activity and haloes' properties and constitutes a constraint to the
semi-analytic model at $z=0$, and that $(ii)$ the cosmological
evolution of DM haloes is well described by the EPS theory. Moreover,
like in the WL03 case, they further assume that $(iii)$ the QSO
activity is triggered by major mergers.

The numerical implementation of the semi-analytic model consists of a
two-step procedure. The first step is aimed at constructing a set of
halo merging histories using the EPS theory. In the EPS formalism,
when one takes a small step $\delta z$ back in time, the number of
progenitors a parent halo of mass $M_0$ at $z=z_0$ fragments into is
\citep{lacey1993}:
\begin{equation}
\frac{dN}{dM}(z=z_0)=\frac{1}{\sqrt{2\pi}}\,\frac{M_0}{M}\,\frac{1}{S^{3/2}}\,
\frac{d\delta_c}{dz}\,\frac{d\sigma_M^2}{dM}\,\delta z\ ,
\label{dn/dmdt}
\end{equation}
where  $S\equiv \sigma_M^2(z)-\sigma_{M_0}^2(z_0)$, $\sigma^2_M(z)$ 
and $\sigma_{M_0}^2(z_0)$ are the linear theory rms density fluctuations smoothed 
with a `top-hat' filter of mass $M$ and $M_0$ at redshifts $z$ and $z_0$,
respectively, and $\delta_c(z)$ is the critical thresholds on 
the linear overdensity for spherical collapse at redshift $z$.
Integrating this function over the range $0<M<M_0$ gives unity: all the mass 
of $M_0$ was in smaller subclumps at an earlier epoch $z>z_0$.  
From Eq.(\ref{dn/dmdt}) we can compute a 
fragmentation probability that, via rejection methods, can be used to
construct a binary merger tree. Implementing a successful Monte Carlo
procedure, however, requires the use of two different numerical
approximations \citep{somerville1999}. First of all, since in a CDM
cosmology the number of haloes diverges as the mass goes to zero, it
is necessary to introduce a cut-off mass, $M_{\rm res}$, that marks
the transition from resolved progenitors (having $M>M_{\rm res}$) to
the accreted mass that accounts for the cumulative contribution of all
mass accreted from unresolved haloes. Secondly, the time-step $\delta
z$ has to be small enough to guarantee a small mean number of
fragments $N_p$ in the range $M_{\rm res}<M<M_0/2$, to avoid multiple
fragmentation.

Once the appropriate choices for $\delta z$ and $M_{\rm res}$ are
made, the binary tree is constructed by using a Monte Carlo procedure
similar to that of \cite{cole2000}, described in VHM.
We have taken $M_{\rm res}=10^{-3}M_0$ at $z=0$ decreasing with
redshift as $(1+z)^{3.5}$.
Finally, for each Monte Carlo realization we have used 820 time-steps
logarithmically spaced between $z=0$ and $z=20$. As shown by VHM,
with this parameter choice our merger tree algorithm not only
conserves the mass, but also reproduces the EPS conditional mass
function at all redshifts.

In the second step of the procedure, we implement a set of analytic
prescriptions and follow the accretion history of massive BHs within
their host haloes to model the QSO activity. The VHM model assumes
that the seed BHs formed with masses of $150 M_\odot$ (note that, as
shown by VHM, the final results are not very sensitive to this choice)
following the collapse of the very rare Pop III stars, in minihaloes
forming at $z=20$ from the density peaks above a $3.5\sigma$
threshold. In the assumed $\Lambda$CDM cosmology this corresponds to
minihaloes with mass $\sim 1.6 \times 10^{7} M_\odot$. Then we assume
that the quasar activity is triggered only by major mergers with
$P>0.1$, a threshold lower than $P>0.25$ adopted by WL03.

\begin{figure}
\includegraphics[width=0.45\textwidth]{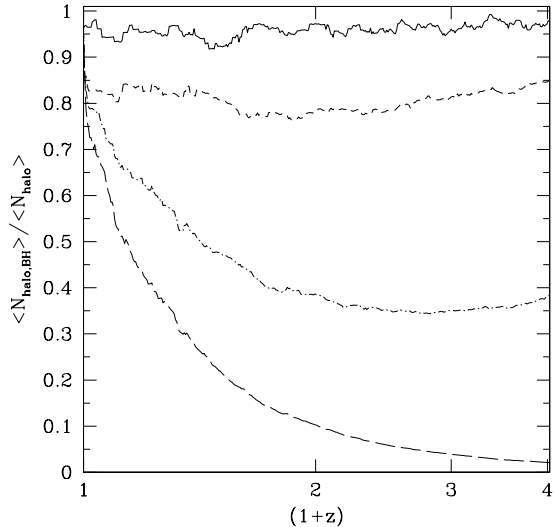}
\caption{Fraction of haloes hosting at least a nuclear 
massive BH vs. redshift. The long-dashed curve shows the occupation
fraction computed by weighting over all branches of the merger trees.
The occupation fraction increases with increasing halo mass:
$M_{\rm halo}>10^{10} M_\odot$ (dot-dashed curve), $M_{\rm halo}>10^{11}
M_\odot$ (short-dashed curve) and $M_{\rm halo}>10^{12} M_\odot$ (solid
curve).}
\label{fig:occfrac}
\end{figure}

Two main features differentiate the VHM model from the WL03 one.
First, the VHM model is naturally biased, as the BH seeds are
associated with high-density peaks in the fluctuations field. Second,
VHM take into account the dynamical evolution of BHs, including strong
gravitational interactions such as the gravitational rocket effect
\citep{merritt2004, volonteri_perna2005}. Such dynamical
interactions can possibly eject BHs at high velocities from the centre
of haloes. The net effect is to contribute to selecting massive haloes
(i.e. those with a large escape velocity) as BH hosts. In all three
scenarios we consider (see below), we have included a treatment of the
`gravitational rocket' effect following \cite{favata2004} and
\cite{merritt2004} (upper limit to the recoil velocity). 
More details can be found in
\cite{volonteri_perna2005}. Fig.~\ref{fig:occfrac} shows the
occupation fraction, i. e. the fraction of haloes hosting nuclear BHs
for haloes of different masses. The occupation fraction of large
haloes ($M_{\rm halo}>10^{12}M_\odot$) is of order unity at all times,
while smaller haloes have a large probability of being deprived of
their central BH.

\begin{figure*}
\includegraphics[width=0.96\textwidth]{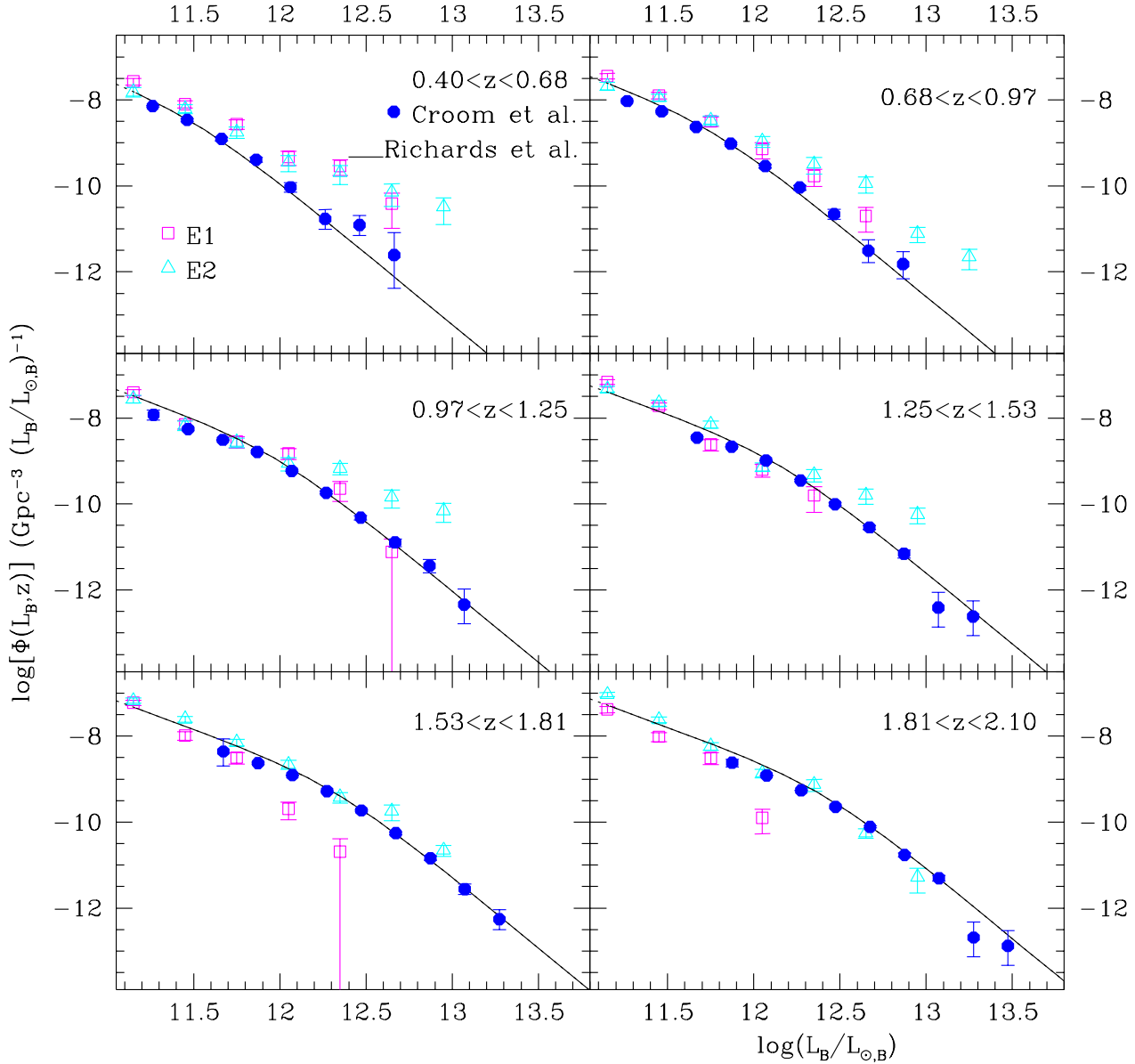}
\caption{
The QSO luminosity function in $B$-band at six different redshifts:
models vs. observations. The filled dots show the 2dF/6dF QSO
luminosity function measured by C04 together with their $1\sigma$
error bars. The thin solid line shows the best fit to the 2SLAQ QSO
luminosity function of R05. The open squares refer to the VHM model
predictions when a fraction $\alpha=7\times 10^{-6}$ of accreted mass
of the merged system total mass is assumed. The open triangles show
the VHM model predictions when a relation between the accreted mass
and the circular velocity of the host halo is assumed.}
\label{fig:lf_sa1}
\end{figure*}

\begin{figure*}
\includegraphics[width=0.96\textwidth]{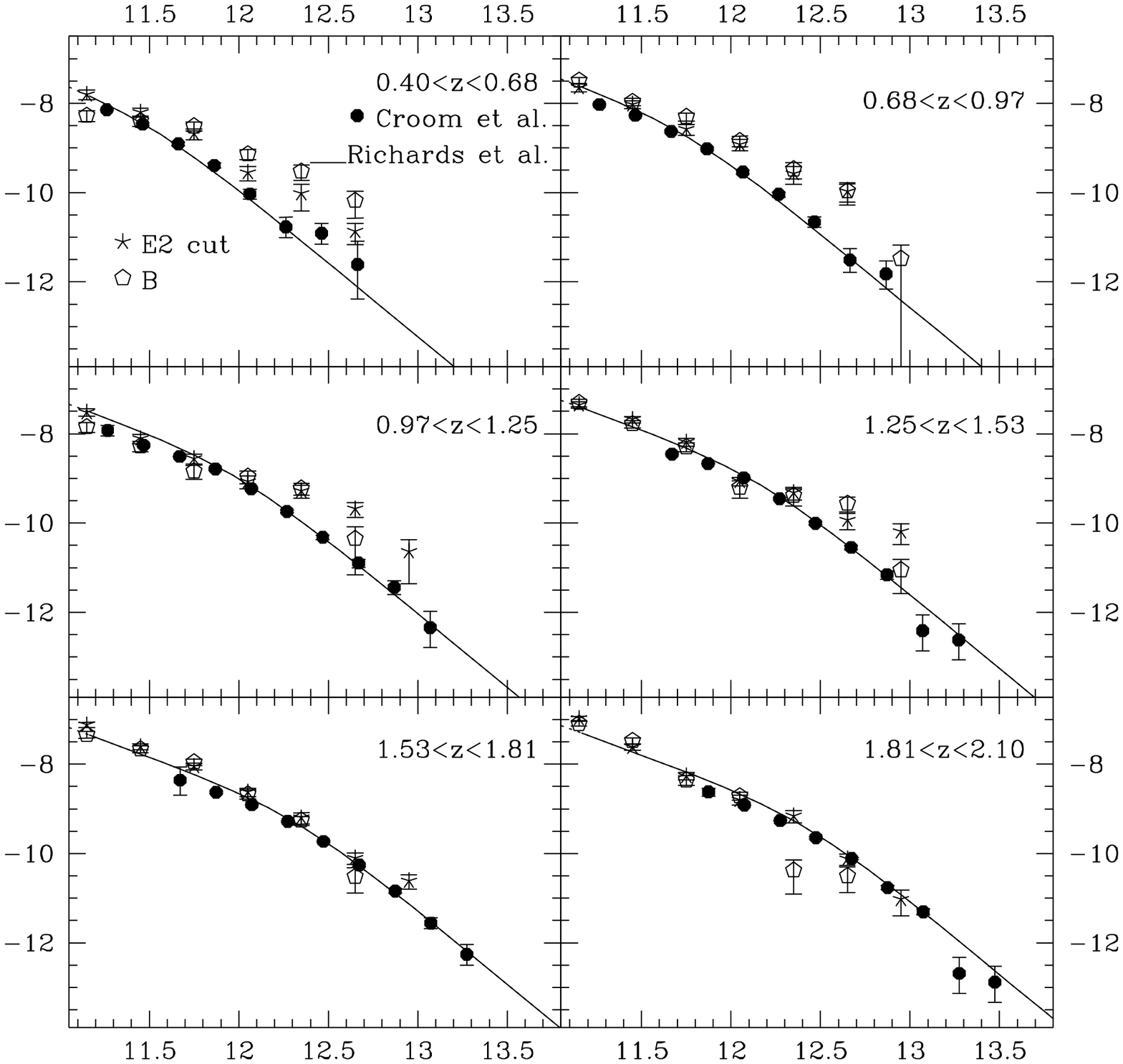}
\caption{
The QSO luminosity function in $B$-band at six different redshifts:
models vs. observations. The filled dots show the 2dF/6dF QSO
luminosity function measured by C04 together with their $1\sigma$
error bars. The thin solid line shows the best fit to the 2SLAQ QSO
luminosity function of R05. The open pentagons refer to the VHM model
predictions including super-critical accretion in high-$z$ haloes. The
asterisks show a model in which the accreted mass scales with the
circular velocity of the host halo, and accretion is suppressed in
haloes with $M_{\rm halo}>10^{13.5}M_\odot$.}
\label{fig:lf_sa2}
\end{figure*}

Following a major merger, the BH at the centre of the massive
progenitor can grow in mass in two ways: $(i)$ after a dynamical
friction time when a bound BH binary system forms at the centre of the
halo, hardens via three body interactions \citep{quinlan1996,
milosavljevic2001} and then rapidly coalesces through the emission of
gravitational waves \citep{peters1964}; $(ii)$ after a dynamical
free-fall time when a significant fraction of the gas falls to the
centre of the merged system \citep{springel2005b, dimatteo2005} and is
accreted on the BH at an appropriate rate. Yet, as shown by
VHM, the first mechanism contributes little to the BH mass accretion
and will be neglected in this work. To implement the second mechanism
we need to specify the prescription for the mass accretion and its
rate. We have explored three different scenarios. The first two
assume that BHs start accreting mass at the Eddington rate
after about one dynamical free-fall timescale from the merger. 
Accretion lasts until a mass
$\Delta M_{\rm accr}$ has been added to the BH, but, as in VHM, we
have inhibited gas accretion in all haloes with $v_c>600 \kms$. 

In the first of the two scenarios, labeled E1, the accreted mass is
proportional to the mass of the available gas and hence to the total
mass of the massive progenitor: $\Delta M_{\rm accr}=\alpha M_{\rm
halo}$. Here, $\alpha=7\times 10^{-6}$ guarantees the normalization of
the $M_{\rm bh}-\sigma_g$ relation at $z=0$, where $\sigma_g$ is the
velocity dispersion of the host galaxy \citep{tremaine2002}, scaling
with the circular velocity of the halo as suggested by
\cite{ferrarese2002}. No feedback is explicitly included in the
semi-analytic scheme. As $M_{\rm halo} \propto v_c^3$, the slope of
the $M_{\rm bh}-\sigma_g$ relation is flatter than the observed one
\citep[but see][]{wyithe2005a}. This scenario is similar in spirit to
the WL03 model, and is meant to compare the clustering properties of
quasars at low redshift to that of their higher redshift
counterparts. \cite{adelberger2005} indeed find that the clustering of
active nuclei at $2\lesssim z \lesssim 3$ points to a $M_{\rm
bh}-M_{\rm halo}$ relation which is independent of redshift.

The second scheme [E2] assumes a scaling relation between the accreted
mass and the circular velocity of the host halo, $\Delta
M_{\rm accr}=k\times v_c^5$, which is normalized {\it a-posteriori} to
reproduce the observed relation between $M_{\rm bh}$ and $v_c$ at $z=0$
\citep{ferrarese2002}. VHM showed that this scenario overestimates 
the optical LF at $z<1$. Here, we assume a linear dependence of $k$
on redshift, as $k=k(z)=0.15(1+z)+0.05$, to account for the decrease
of the gas available to fuel BHs. The above relation is totally
empirical and was determined iteratively in order to get a better fit
to the LF, without underestimating the local SBH mass density.

The last prescription for mass accretion, labeled B, assumes an early
stage of super-critical accretion during which the central BH accretes
mass at a rate that can be estimated by the Bondi-Hoyle formula
\citep{bondi1944}. This model applies to metal-free haloes, therefore
we assume that by $z=12$ the interstellar medium has been enriched,
and we inhibit super-critical accretion rates. When the super-critical
phase ends, accretion proceeds in subsequent episodes as in model E2.
This possibility has been recently advocated by
\cite{volonteri_rees2005} to reconcile a hierarchical evolution with
the existence of QSO at $z\sim 6$, hosting SBHs with masses $\sim 10^9
M_\odot$.

The end product of our semi-analytic models is a set of merging and
accretion histories for 220 parent haloes with masses in the range
$(1.43\times10^{11}M_\odot , 10^{15}M_\odot)$. When active,
i.e. during the period of mass accretion, the QSO shines with a
$B$-band luminosity of $(L_B/L_\odot)=M_{\rm bh}\times10^{3.46}
M_\odot$, obtained under the assumptions that the rest mass is
converted to radiation with a 10 per cent efficiency and that only a
fraction $f_{\rm B}=0.08$ of the bolometric power is radiated in the
blue band.
Finally quasars in the model outputs are selected according to the
2dF/6dF criteria.

The model QSO LF at different redshifts has been computed by
evaluating the number density of active QSOs in each luminosity bin in
redshift intervals centred on an effective redshift $z_{\rm eff}$. In
practice we have counted the number of active QSOs in the redshift and
luminosity bins in all merger trees, each of them weighted by the
number density of their parent haloes at $z=0$ [evaluated using the
\cite{sheth1999} formula]; the result has been normalized using 
the number of time-steps in the redshift interval and the number of
merger trees considered. Associate uncertainties have been computed
by assuming Poisson statistics.

The biasing function $b(z)$ at the three redshifts considered by PMN
has been estimated by using the following equation
\begin{equation}
b(z)=\frac{\displaystyle\int_{0}^{+\infty}b(M_{\rm halo},z)
\Psi(M_{\rm halo}(L_B>L_{{\rm min},B}),z)dM_{\rm halo}}
{\displaystyle\int_{0}^{+\infty}\Psi(M_{\rm halo}
(L_B>L_{{\rm min},B}),z)dM_{\rm halo}}\ ,
\end{equation}
where $\Psi(M,z)$ is the mass function of the haloes hosting QSOs with
luminosities larger than the selection thresholds of PMN ($L_{B,{\rm
min}}/L_{\odot,B}=\{1.5\times10^{11},3.9\times10^{11},
6.6\times10^{11}\}$), and $b(M_{\rm halo},z)$ is the bias function of
haloes computed following \cite{sheth1999}.

Model predictions also allow a straightforward evaluation of the mean
halo occupation number, i.e. the average number of active QSOs hosted
in haloes with mass between $M_{\rm halo}$ and $M_{\rm halo}+dM_{\rm halo}$:
\begin{equation}
N_{\rm QSO}(M_{halo})=\frac{\Psi(M_{\rm halo})}{dN/dM_{\rm halo}}\ ,
\end{equation}
where $\Psi(M_{\rm halo})$ is the mass function of haloes hosting active
quasars in the halo merger trees and $dN/dM_{\rm halo}$ is the
\cite{sheth1999} halo mass function. Model uncertainties have been
evaluated from Poisson errors associated to $\Psi(M_{\rm halo})$.

\subsubsection{VHM model vs. observations}
\label{sec:sa_lf}

\begin{figure}
\includegraphics[width=0.45\textwidth]{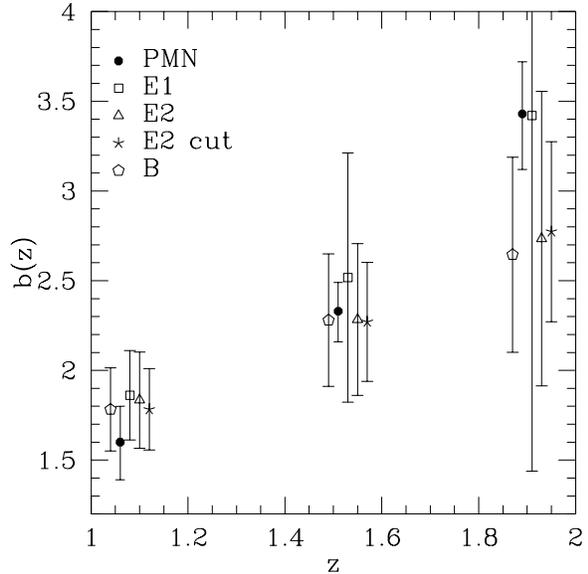}
\caption{
The mean QSO-to-mass biasing parameter, $b(z_{\rm eff})$ estimated at
three effective redshifts $z=1.06$, $z=1.51$ and $z=1.89$: models
vs. observations. The filled circles show the mean biasing of 2dF/6dF
quasars measured by PMN. Symbols are as in Figs.~\ref{fig:lf_sa1} and
\ref{fig:lf_sa2}.}
\label{fig:bias_sa}
\end{figure}

\begin{figure}
\includegraphics[width=0.45\textwidth]{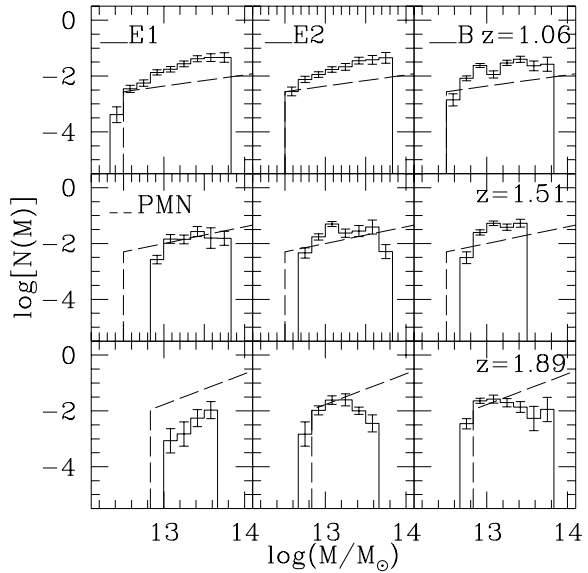}
\caption{
The mean halo occupation number of QSO at three effective redshifts
$z=1.06$, $z=1.51$ and $z=1.89$ (from top to bottom). Histograms
represent model predictions, while the dashed line shows the mean halo
occupation number proposed by PMN and consistent with 2dF/6dF QSO
data. Left panels: model E1. Central panels: model E2. Right panels: model B.}
\label{fig:nm_sa}
\end{figure}

\begin{figure}
\includegraphics[width=0.45\textwidth]{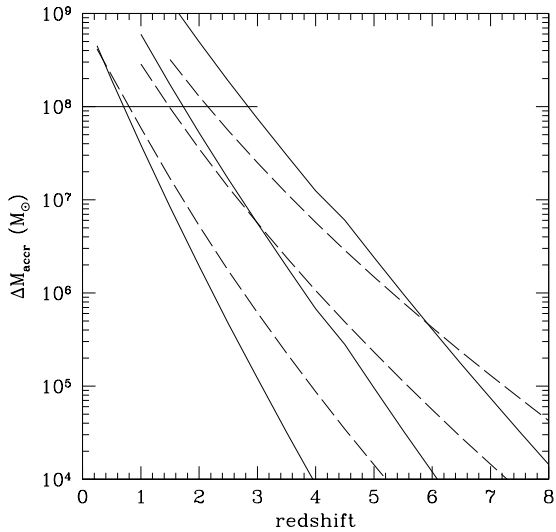}
\caption{
Accreted mass as a function of redshift for different halo
masses. From top to bottom, curves are for $2.5\sigma$, $2\sigma$ and
$1.5\sigma$ peak haloes. Dashed curves: model E1. Solid curves: model
E2. The horizontal line shows the BH mass corresponding to $L_{B,{\rm
min}}$ at $z=1.89$ of PMN, assuming Eddington accretion rate.}
\label{fig:dmacc}
\end{figure}

In Figs.~\ref{fig:lf_sa1} and \ref{fig:lf_sa2} we compare our model LF
with the observations. The 6dF/2dF and the 2SLAQ QSO LFs are plotted
using the same symbols as in Fig.~\ref{fig:lf}. The LFs predicted by
our semi-analytic models are plotted with the three different symbols
indicated in the plot. At low redshifts all models reproduce the
faint end of the LF fairly well.
However, they systematically overpredict the number of bright QSOs:
this indicates that having inhibited gas accretion in haloes with
$v_c>600 \kms$ has little impact on our results. Indeed, this
circular velocity is significantly larger than that associated to the
mass threshold $M^{*}_{\rm halo}\sim10^{13.5} M_{\odot} $ adopted in
the WL03K model. Imposing a similar mass threshold in the VHM model
(model E2$_{\rm cut}$), in place of the $v_c>600 \kms$ cut-off, 
we find a much better agreement between the
model predictions and the observed LF at all $z<1.5$. Finally, we note
that model E1 fails to produce QSOs brighter than $ 10^{12}
L_{\odot,B}$.

As shown in Fig.~\ref{fig:bias_sa}, all semi-analytic models explored
match the observed biasing function out to $z\sim2 $. We notice that
the three semi-analytic models explored predict a very similar biasing
function. As pointed out by \cite{wyithe2005a}, the evolution of
clustering is slightly faster when the BH mass scales with the halo
mass (model E1) rather than the circular velocity (model E2), although
the difference is of little significance given the large scatter in
the model predictions.

The reason of the difference can be understood by looking at the
histograms plotted in Fig.~\ref{fig:nm_sa} representing the mean halo
occupation number, $N_{\rm QSO}(M_{\rm halo})$, of the various models.
At $z=1.89$ model E1 exhibits a steeper dependence on halo mass
compared to the other cases meaning that QSOs are preferentially found
in massive haloes with a high degree of spatial clustering. This
result derives from the fact that the accretion scheme of model E1 is
more efficient at very high redshift ($z>6$) than that of model E2,
due to the different scaling of $\Delta M_{\rm accr}$ with redshift.
Fig.~\ref{fig:dmacc} exemplifies this effect. On one hand, $\Delta
M_{\rm accr}$ is a steeper function of redshift in model E2, implying
that massive BHs accrete more mass in every accretion episode, thus
leading to a longer duty-cycle, and a larger occupation number, in
general. On the other hand, at redshift $z=1.89$, only quasars above
$L_{B,{\rm min}}$ are selected. In model E2 the $2\sigma$-peak haloes
contain BHs massive enough to be above threshold, while for model E1 a
slightly more massive halo is needed. Consequently this enhances the
bias. At lower redshifts this effect becomes progressively less
important. This result interestingly agrees with that of
\cite{adelberger2005}: at $z>2$ SMBH masses correlate with the
halo mass instead of with velocity dispersion, or circular velocity.
We can speculate that there might be a transition in the interaction
between SMBHs and their hosts, which switches on at $z\simeq 2$: at
higher redshifts the SMBH mass scales with the halo mass, at lower
redshifts with its velocity. As shown in Fig.~\ref{fig:dmacc}, the
accreted mass in a given episode is larger in model E1 than in model
E2 for haloes representing density peaks below $2\sigma$ at
$z>3$. Model E1, therefore, implies an earlier growth for SMBHs.

On the contrary, model B, in which massive BHs accrete mass with very
high efficiency at high redshifts preferentially populate smaller
haloes. Massive BHs in more massive haloes have already grown to
masses close to the $M_{\rm bh}-\sigma_g$ threshold. At $z=1.06$ all
models, especially E1, predicts a number of QSOs in haloes which is
systematically larger than the one inferred from PMN, especially in
high-mass haloes. This discrepancy, which is marginally significant, considering the
errors estimated by PMN, reflects the fact that
semi-analytic models overestimate the optical LF of bright quasars at
low redshifts, as shown in the top panels of Figs.~\ref{fig:lf_sa1} and
\ref{fig:lf_sa2}. 
On the other hand, at high redshifts all models
predict a halo occupation number that is systematically smaller than
the PMN one, which again reflects the fact that our model LFs slightly
underestimate the observed one at high redshifts
(Figs.~\ref{fig:lf_sa1} and \ref{fig:lf_sa2}, bottom panels). This
effect is particularly evident for model E1, which, as we have noticed
before, predicts no QSOs with $ L> 10^{12} L_{\odot,B}$ at $z>1.25$.

\section{Discussion and Conclusions}
\label{sec:sa_lf1}

\begin{figure*}
\includegraphics[width=0.96\textwidth]{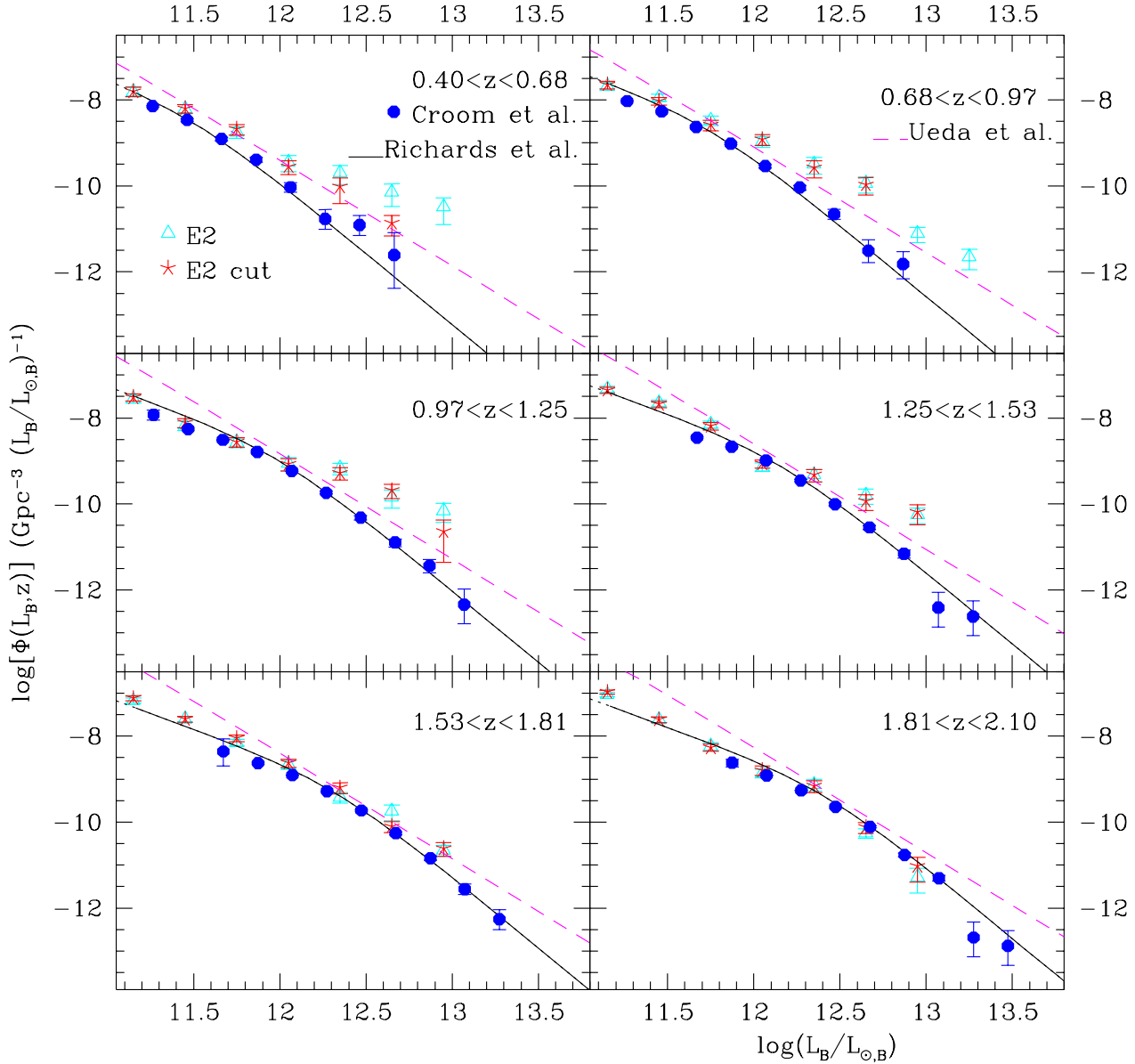}
\caption{
Comparison of the model
predictions to the observed optical QSO LF (filled circles: C04; thin
line: R05) and the hard X-ray LF
\protect\citep{ueda2003}, converted into $B$-band (see text for more details).
Symbols are as in Figs.~\ref{fig:lf_sa1} and \ref{fig:lf_sa2}.}
\label{fig:LFX}
\end{figure*}

In this work, we have carried out for the first time a systematic study aimed at
assessing how well, within the framework of hierarchical build up
of supermassive black holes and massive halos hosts, analytic and 
semi-analytic techniques perform in reproducing the QSO luminosity 
function and spatial distribution at very low redshifts, bound to reproduce the 
relation between $M_{\rm bh}$ and $M_{\rm halo}$ 
\citep{ferrarese2002}.
We have found that only minor, physically-motivated modifications 
to  standard semi-analytic techniques are required to match observations
at redshifts as small as 0.5.
More profound modifications seem to be required for a 
successful modeling of the very local QSO population.

Previous studies, like those performed by WL02, WL03 and VHM, have
shown that analytic and semi-analytic models well reproduce the
observed QSO LF at high redshifts. At low redshifts, however, these
methods systematically overpredict the number density of bright
objects; a mismatch that becomes increasingly large when decreasing
the redshift. To improve the fit to the data we have proposed two
simple modifications to the original WL02, WL03 and VHM models. The
first possibility is to assume that the major merger threshold
required to trigger the QSO activity depends on the mass of the
hosting halo. This reduces the number density of bright QSOs that can
only be activated by rare merging events between large haloes of
similar masses. A second possible modification is physically
motivated by the fact that gas accretion is expected to be inefficient
within large haloes due to the high temperature of the baryons: this
can inhibit accretion in haloes larger than $\sim10^{13.5} M_{\odot}$,
hence reducing the abundance of bright objects. Both prescriptions
significantly improve the fit to the bright end of the QSO LF,
especially when implemented within the analytic framework. Yet,
significant discrepancies still remain, especially at very low
redshifts, that could be possibly eliminated by including two more
factors that are missing from the models.

First, when comparing our model predictions to the observed $B$-band
QSO LF, we have implicitly ignored the presence of a substantial
population of (optically) obscured, luminous AGN at low redshifts,
whose existence, instead, is suggested by {\it Chandra} results
\citep[see, e.g.,][]{barger2001,rosati2002}. In our modeling we have
not included any correction for type-II quasars, which appear to
comprise between 30 per cent \citep{lafranca2005}, and $\sim 80$ per
cent \citep{brown2005, franceschini2005} of the quasar population at
$0.5<z<2$. However, our model successfully reproduces the QSO LF in
the hard X-ray band. Fig.~\ref{fig:LFX} shows a comparison between the
model predictions and the [2-10 keV] LF obtained by \cite{ueda2003}.
The conversion from the hard-X band luminosity into a blue-band
estimate has been done by assuming the scaling proposed by
\cite{marconi2004}. We also compared the integrated mass density as a
function of redshift to the redshift dependent integral of luminosity
density of quasars \citep{marconi2004}, including the X-ray LF
\citep{ueda2003} and a correction of a factor 2 to account for missing
Compton-thick quasars \citep{brown2005}. We found a very good
agreement over the whole redshift range probed by the LFs ($0<z<3$).
We therefore ascribe the discrepancy between model predictions and
observations at optical wavelengths mainly to selection effects.
 
Second, \cite{merloni2003} and \cite{merloni2004} have shown that
low-redshift AGN are probably accreting inefficiently, i.e. both at an
accretion rate much smaller than the Eddington rate and with a low
radiative efficiency. Evidence for a BH powering mechanism less
efficient at low redshift is also provided by the fact that local
bright QSOs seem to be hosted in early-type galaxies that show no sign
of recent merging events like disturbed morphology or recent star
formation episodes \citep[see, e.g.,][ and references therein]
{grogin2005, grazian2004, dunlop2003}. These considerations suggest that a
successful model for describing the evolution of QSO luminosity should
include both a prescription for a frequency-dependent galaxy
obscuration and a more sophisticated mechanism for the QSO activity in
which the BH accretion rate and the QSO duty-cycle might depend on
halo masses and merger parameters, as suggested by recent numerical
experiments \citep{dimatteo2005}.

Both analytic and semi-analytic models predict a moderate degree of
QSO clustering at low redshift, consistent with the observations. At
higher redshifts ($z\approx 2$) the QSO biasing predicted by the
analytic models appears to be significantly smaller than that of
2QZ/6QZ quasars. The only way for reproducing the observed degree of
clustering is to increase the normalization constant $\epsilon$ in the
$M_{\rm bh}-v_c^{\gamma}$ relation \citep{wyithe2005a}, which,
however, would overpredict the QSO number density in the local
universe.

On the contrary, the redshift evolution predicted by semi-analytic
schemes is significantly faster and matches observations out to
$z\sim2$. The high degree of clustering predicted by the
semi-analytic VHM model does not derive from having placed the first
seed BHs in correspondence of high-$\sigma$ overdensity peaks of the
mass density field. The adopted threshold, $3.5\sigma$ at $z=20$, in
fact corresponds to having at least one seed massive BH in haloes with
$M_{\rm halo}\simeq10^{11} M_\odot$ at $z=0$, which host massive BHs
with masses well below that sampled by the optical LF of quasars.
Moreover, dynamical effects such as the gravitational rocket
\citep{volonteri_perna2005} can possibly eject BHs and thus lower the
occupation fraction only in haloes with $M_{\rm halo}<10^{12} M_\odot$
that host BHs too faint to be included in the range probed by the
optical LF. Placing seed BHs in correspondence of even higher peaks
would certainly increase the biasing of the QSOs without modifying
their luminosity function at $z>0.5$, provided that the major merger
threshold is changed accordingly (VHM). In this case, however, it
would be difficult to explain the presence of SBHs in galaxies like
the Milky Way or smaller, and, in general, AGN harboured in dwarf
galaxies \citep{barth2005}, which anyway are not sampled by the quasar
LF at $z>0.4$.
The large values of bias in the semi-analytic models have a different
explanation: it derives from the lack of a deterministic relation
between the DM halo masses and the QSO luminosities at a given time.
Indeed, a finite time is required to accrete a mass $\Delta M_{\rm
accr}$ to the central BH. During the accretion phase, the BH is
smaller than predicted by the simple scaling relations with halo
masses (i.e. $M_{\rm bh}-\sigma_g$ or $M_{\rm bh}-M_{\rm halo}$). This
means that, on average, the hosting halo of a quasar of a given
luminosity is larger in the semi-analytic scheme than in the
analytic models, the masses being the same only at the very end of
the accretion episode. Consequently, the bias is enhanced in the
semi-analytic model even if the LF looks similar.

In our analysis we have shown that simple models in which QSO activity
is triggered by halo mergers within the framework of hierarchical
build-up of cosmic structures can quantitatively describe the observed
evolution of the quasar number counts and luminosity at all but very
low redshifts, provided that some mechanisms are advocated to inhibit
accretion within massive haloes hosting bright quasars. Semi-analytic
schemes, that naturally account for the finite accretion time, are
also capable of reproducing the observed QSO clustering at $z<2$.
These results suggest that the QSO evolution can probably be explained
within the hierarchical scenario of structure formation in which the
QSO activity mainly depends on the masses of their host haloes. More
realistic models, however, should also account for the various halo
properties including those of the baryons and for possible
environmental effects. One possibility, which is most easily
implemented within a semi-analytic scheme, is to include absorption
effects and modify the accretion scheme accordingly using the outcomes
of numerical hydrodynamical experiments. With this respect, we plan
to modify the simple accretion scheme used in the VHM model following
the recent results of \cite{hopkins2005} who found that active QSOs
are heavily obscured for most of their time but during their peak of
activity when the surrounding dust is blown away and the QSOs shine
prominently for a short period ($\sim 10^7$ yr).

It is worth stressing that the hierarchical models for QSO evolution
and the possible modifications discussed so far rely on two important
assumptions which have recently been cast into doubt. First of all,
the models considered in this work assume a simple relation between
QSO activity and the mass of its hosting halo. However,
\cite{wyithe2005b} pointed out that the tight relation between the BH
mass and the velocity dispersion of the spheroid implies that it is the
spheroid rather than the halo which determines the growth of the SBHs
and the subsequent QSO activity. As a consequence, the observed
correlation between the halo and BH masses should not be regarded as
fundamental as it merely reflects the fact that massive haloes
preferentially host bulges with large velocity dispersions. This would
imply that QSO activity, closely related to the evolution of bulges,
should be studied using the more sophisticated models for galaxy
formation and evolution. The second and more important issue is
related to the results of the recent Millennium Simulation
\citep{springel2005}. The analyses performed by \cite{gao2005} and by
\cite{harker2005} have shown that the spatial correlation properties
and the formation epochs of the haloes depend on the local
overdensity. This effect, which is particularly evident for
galaxy-sized haloes, contradicts one of the basic assumption of the
EPS theory which we have used to construct the halo merger history and
to describe their biasing function. It is not clear, however, how
serious the implications are for galaxy/QSO formation models and for
halo models of clustering. In case they are and in absence of a
generalized EPS theory capable of accounting for environmental effects
\citep[see, however,][]{abbas2005,shen2005}, then the only way out
would be that of resorting to halo merger histories extracted from
numerical experiments that implicitly account for environmental
dependencies \citep[see, e.g.][]{lemson1999}. We will pursue this
strategy in a subsequent work.

\section*{Acknowledgements}
FM thanks the Institute of Astronomy, University of Cambridge, 
for the kind hospitality. MV acknowledges the warm and unforgettable 
hospitality at the Universit\`a degli Studi ``Roma Tre''.
We are grateful to Francesco Pace for useful discussions.

\bibliographystyle{mn2e}
\bibliography{master_qso}

\label{lastpage}
\end{document}